\documentclass[a4paper,fleqn]{cas-sc}
\usepackage{ulem}
\usepackage{lineno}
\usepackage[numbers]{natbib}
\usepackage[]{natbib}
\usepackage{caption}
\usepackage{subcaption}
\usepackage{xr}
\usepackage{pifont}
\def\tsc#1{\csdef{#1}{\textsc{\lowercase{#1}}\xspace}}
\tsc{WGM}
\tsc{QE}
\tsc{EP}
\tsc{PMS}
\tsc{BEC}
\tsc{DE}

\makeatletter
\newcommand*{\addFileDependency}[1]{
  \typeout{(#1)}
  \@addtofilelist{#1}
  \IfFileExists{#1}{}{\typeout{No file #1.}}
}
\makeatother

\newcommand*{\myexternaldocument}[1]{%
    \externaldocument{#1}%
    \addFileDependency{#1.tex}%
    \addFileDependency{#1.aux}%
}

\myexternaldocument{SI}

\begin{document}
\let\WriteBookmarks\relax
\def\floatpagepagefraction{1}
\def\textpagefraction{.001}
\shorttitle{Interfacial self-assembly of chitosan / surfactants}
\shortauthors{R Chachanidze et~al.}

\title [mode = title]{Structural characterization of the interfacial self-assembly of chitosan with oppositely charged surfactant}                      
\author[1]{Revaz Chachanidze}\cormark[1]
\author[1,2]{Kaili Xie}
\author[1]{Hanna Massaad}	
\author[1]{Denis Roux}
\author[1,3]{Marc Leonetti}
\author[1]{Cl\'ement de Loubens}[orcid=0000-0002-4988-9168]

\address[1]{Univ. Grenoble Alpes, CNRS, Grenoble INP, LRP, 38000 Grenoble, France}
\address[2]{Univ. Bordeaux, CNRS LOMA UMR 5798, Talence F-33405, France
}
\address[3]{Univ. Aix-Marseille,  CNRS, CINaM, Marseille, France
}
\cortext[cor1]{Corresponding authors:  revaz.chachanidze@univ-grenoble-alpes.fr}

\begin{abstract}
Controlling the assembly of polyelectrolytes and surfactant at liquid-liquid interfaces offers new ways to fabricate soft materials with specific physical properties. However, little is known of the relationships between the kinetics of interfacial assembly, structural and rheological properties of such interfaces.
We studied the kinetics at water-oil interface of the assembly of a positively charged biopolymer, chitosan, with an anionic fatty acid using a multi-scale approach. The growth kinetics of the membrane was followed by interfacial rheometry and space- and time- resolved dynamic light scattering.
This set of techniques revealed that the interfacial complexation was a multi-step process. At short time-scale, the interface was fluid and made of heterogeneous patches. At a ‘gelation’ time, the surface elastic modulus and the  correlation between speckles increased sharply meaning that the patches percolated. Confocal and electron microscopy confirmed this picture, and revealed that the basic brick of the membrane was sub-micrometric aggregates of chitosan / fatty acid. 

\end{abstract}

\begin{graphicalabstract}
\includegraphics[scale=0.5]{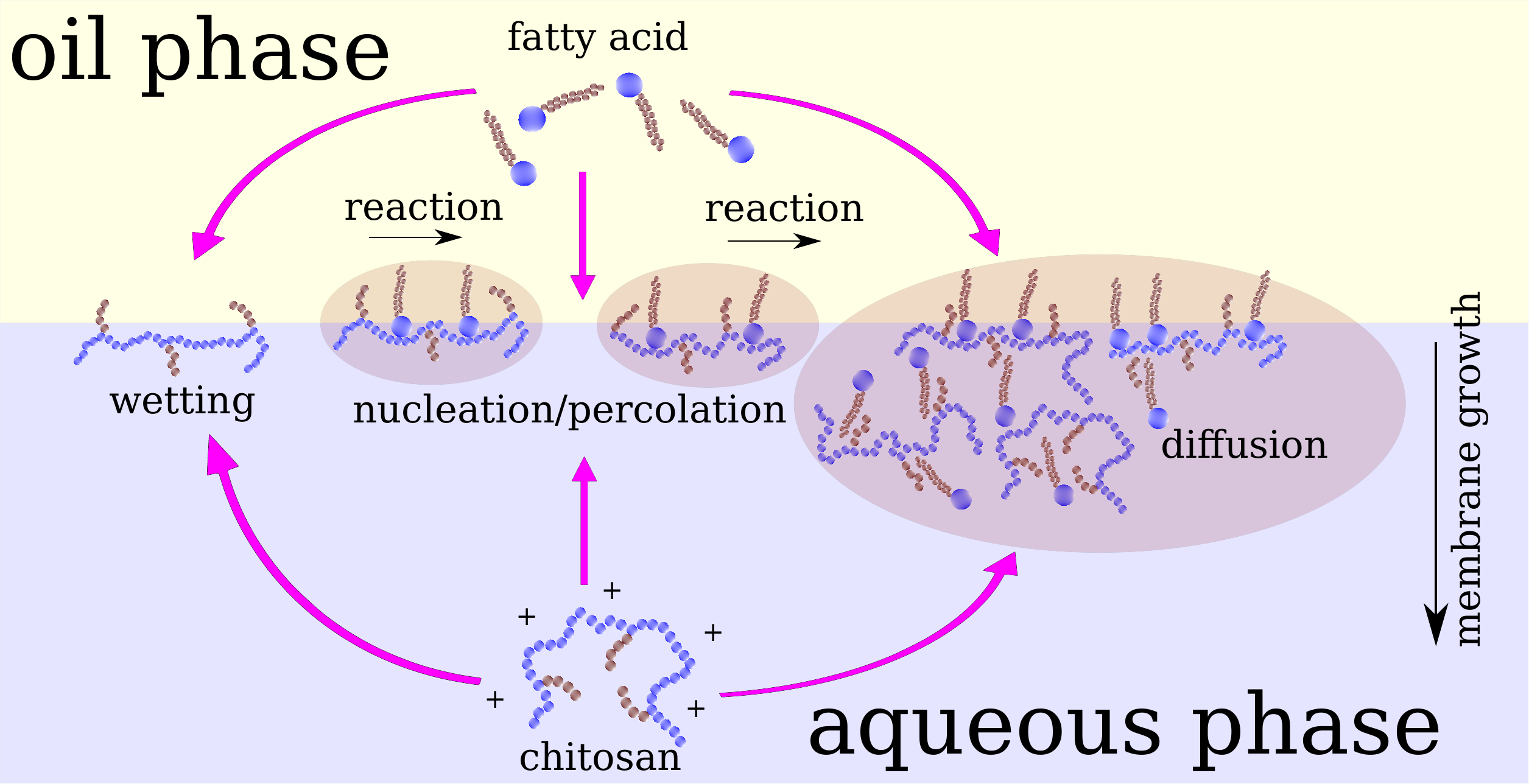}
\end{graphicalabstract}

\begin{highlights}
\item A multiscale study of the interfacial polymerization was conducted.
\item The forming interfacial membrane was probed noninvasively  with a space- and time- resolved DLS approach.. 
\item The results show spatial and temporal heterogeneities of membrane formation.  
\end{highlights}


\maketitle

\linenumbers
\section{Introduction}
Since pioneering observations by Ramdsen \citep{ramsden1904separation} and Pickering \citep{pickering1907cxciv} regarding the stabilization of emulsions and foams by colloidal particles trapped at interface, the interfacial assembly of colloids has seen growing interest from scientific communities.  It opens the way to the fabrication of materials with specific physical properties  such as films, capsules or structured liquids by using interfaces as scaffolds \citep{forth2019building} as well as understanding some physiological functions \cite{bertsch2021physiological}. These materials can be produced by droplet formation \citep{amstad2017capsules} or 3D-printing of liquid-liquid interfaces \citep{xu2020interfacial}. The main driving mechanism behind the self-assembly of colloids at interface is the process of minimization of interfacial energy, which can be tuned by an external stimulus \citep{maestro2019tailoring} or by controlling the interactions between the particles \citep{israelachvili2015intermolecular}. As a result of this assembly, the interface can have a solid-like or liquid-like behaviour. One striking example is the possibility to design mechanically pH-responsive and self-healing microcapsules by interfacial assembly of polymer-polymer coacervates \citep{de2017one}, which open the way to \textit{in-situ} reconfigurable structured liquid interfaces. The rational design of self-assembled membranes with optimal properties  requires an understanding of the interplay between the properties of the colloids, the kinetics of assembly and the structure of the interface.

The relation between the kinetics of interfacial assembly and the resulting physical properties such as membrane thickness or interfacial rheology has been studied for various systems, i.e. nanoparticles, polymers and polyelectrolytes. For interfaces covered by model nanoparticles \citep{thijssen2017interfacial}, the structure of the interface changes with the increasing surface coverage, from a fractal network of aggregates to a heterogeneous structure with voids, to a gel with dense clusters and eventually a densely-packed system \citep{reynaert2007interfacial,masschaele2009direct}. Consequently, viscoelasticity and yield points of these interfaces are controlled by the surface coverage, interparticle interactions and external field forces \citep{masschaele2009direct, cui2013stabilizing}.

Another possible approach stabilising interfaces is via formation of an interfacial complex whereby two oppositely charged polyelectrolytes are dissolved in different immiscible phases \citep{grigoriev2008new,monteillet2013charge,kaufman2014single,kim2015one}. Upon contact, polycations and polyanions diffuse spontaneously towards the interface and form a membrane or a coacervate by electrostatic interactions.  Monteillet \textit{et al.} \cite{monteillet2013charge} studied the kinetics of assembly of polyelectrolytes at water-oil  interfaces at macroscopic scales. They showed that the assembly was a two-stages process: a fast diffusion limited adsorption process which was followed by a much slower logarithmic process. The latter should  result from the hindered interpenetration of the two oppositely charged polymers, such as coacervation in the bulk. Moreover, self-consistent field analysis carried-out by the same group of authors suggested that the 
coacervate film should be heterogeneous \citep{monteillet2017complex}.  H-bond acceptor and
donor polymers have also been used to cover water-oil interfaces by interfacial complexation of both polymers \citep{de2017one}. For these systems, the elasticity is controlled by the type and strength of physical interactions \citep{le2015interplay}. Dupré de Baubigny \textit{et al.}  \citep{de2020entropically} investigated the kinetics of membrane growth on long time scales (>\,1,000\,s) and identified a diffusion limited process. However, the authors were surprised to observe that the process was faster when polymer molar mass increased. They related this observation to the description of the structure of the membrane as a gel-like porous network, with a pore size much smaller than the radius of the diffusing polymer chains. As a result, the diffusion process should be hindered by the enthropic barrier.  

There is a growing interest in the systems formed through a complexation between polyelectrolytes  and oppositely charged surfactant.   Such a system has proved very useful in stabilising emulsions \cite{mun2006effect, thanasukarn2006utilization},   microcapsule fabrication  \cite{gunes2011tuneable,kuroiwa2021biocompatible,xie2017interfacial,maleki2021membrane} and  liposome coatings \cite{mady2009biophysical}. The driving force  is the entropic gain due to the release of the counter-ions and water molecules \cite{chiappisi2015co}.  In this paper, we study such process  by using a model system namely complexation between a water-soluble  positively charged biopolyelectrolyte, chitosan, with an oil-soluble anionic surfactant. Chitosan is highly positively charged in acidic medium due to the protonation of its amino groups (NH3$^+$). The electrostatic interactions between positive NH3$^+$ groups of chitosan and oppositely charged surfactants is controlled by the degree of acetylation of chitosan and pH \cite{rinaudo2006chitin,chiappisi2015co}. In the case of vesicles, chitosan interacts with the phospholipds bi-layer and its coverage can be strongly heterogeneous with the formation of holes \cite{mertins2013insightsI,mertins2013insightsII}. In the case of microcapsules, the membrane is formed by the complexation of chitosan with a fatty acid at the water-oil interface. The elasticity of the membrane increases with time and the concentration of short chain fatty acid \cite{gunes2011tuneable, xie2017interfacial}.

All these results obtained with different systems highlights the importance of structural characterization at different scales, as the interfacial coverage can be strongly heterogeneous.  Moreover, given the difficulties associated with comparing the results, it is important to combine different methodologies using various interfacial characterization tools \cite{biviano2021viscoelastic}.

The aim of our work is to describe the kinetics of the assembly of a polyelectrolyte with an oppositely charged surfactant at water-oil interface using a multiscale approach. In our study, chitosan, a water soluble cationic polymer, was used to  form a complex with oil-soluble anionic  phosphatidic acids at water-oil interface. This system has been used for microcapsule production \citep{gunes2011tuneable,xie2017interfacial, maleki2021membrane}. The interest of the model system for the present study, lies in the fact that its kinetics is relatively slow to study the different stages of the assembly. We characterized the kinetics of assembly at macroscopic scales by interfacial rheometer to follow the "gelation" of the interface with measurement of the velocity field of the interface. At nanometric scales, we developed space- and time- resolved  dynamic light scattering (DLS) to characterize the changes in the heterogeneities of the interfaces, which was complemented by confocal and scanning electron microscopies (CSM and SEM). Lastly, this approach allowed us to relate the structure of the forming film with its rheological properties. We discuss also the analogies between this system of polyelectrolyte / surfactant with others stabilized interfaces systems. 

\begin{figure}[tb!]
	\centering
    \includegraphics[scale=0.5]{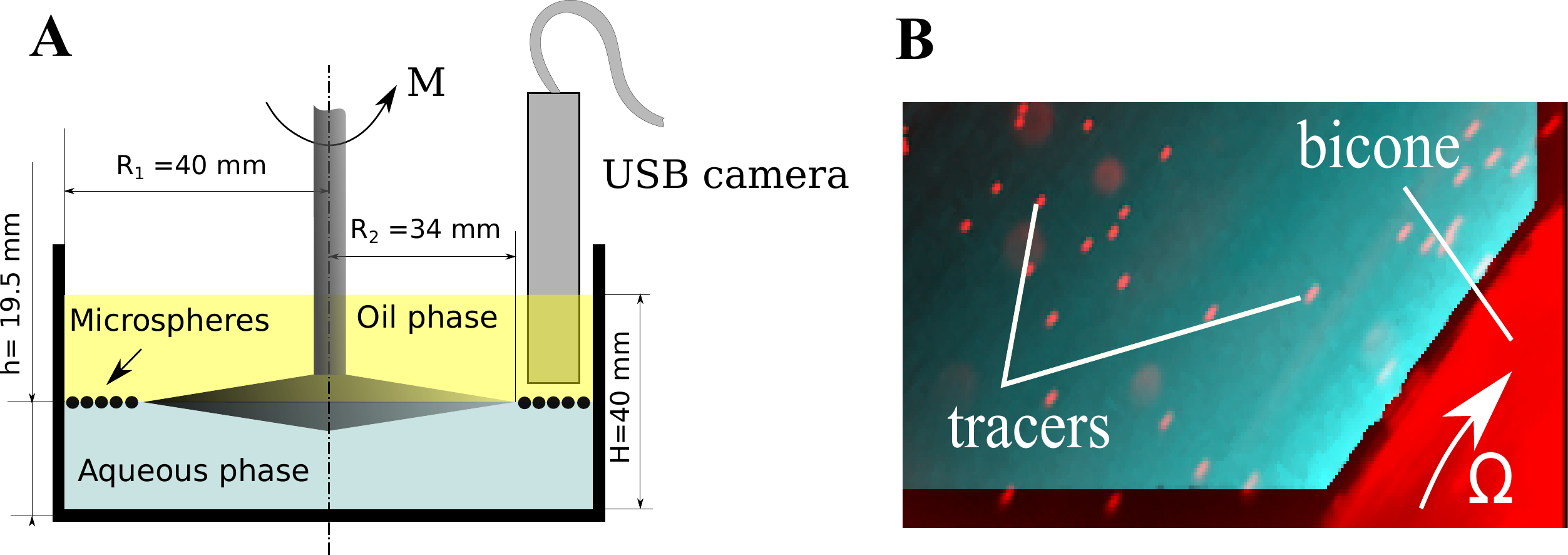}
    \caption{Inerfacial rheometry by means of IRS. (A) Schematic representation of the bicone  rheological cell used to probe the interfacial  properties of chitosan / PFacid membrane. The interface was seeded with microparticles in order to visualize the velocity field of the interface with an immersed camera. (B) PTV at the water-oil interface, the color gradient shows the average velocity of tracers decreasing further when moving away from bicone.}
	\label{FIG:interfacial-rheometry}
\end{figure}

\begin{figure}[tb!]
	\centering
    \includegraphics[scale=0.38]{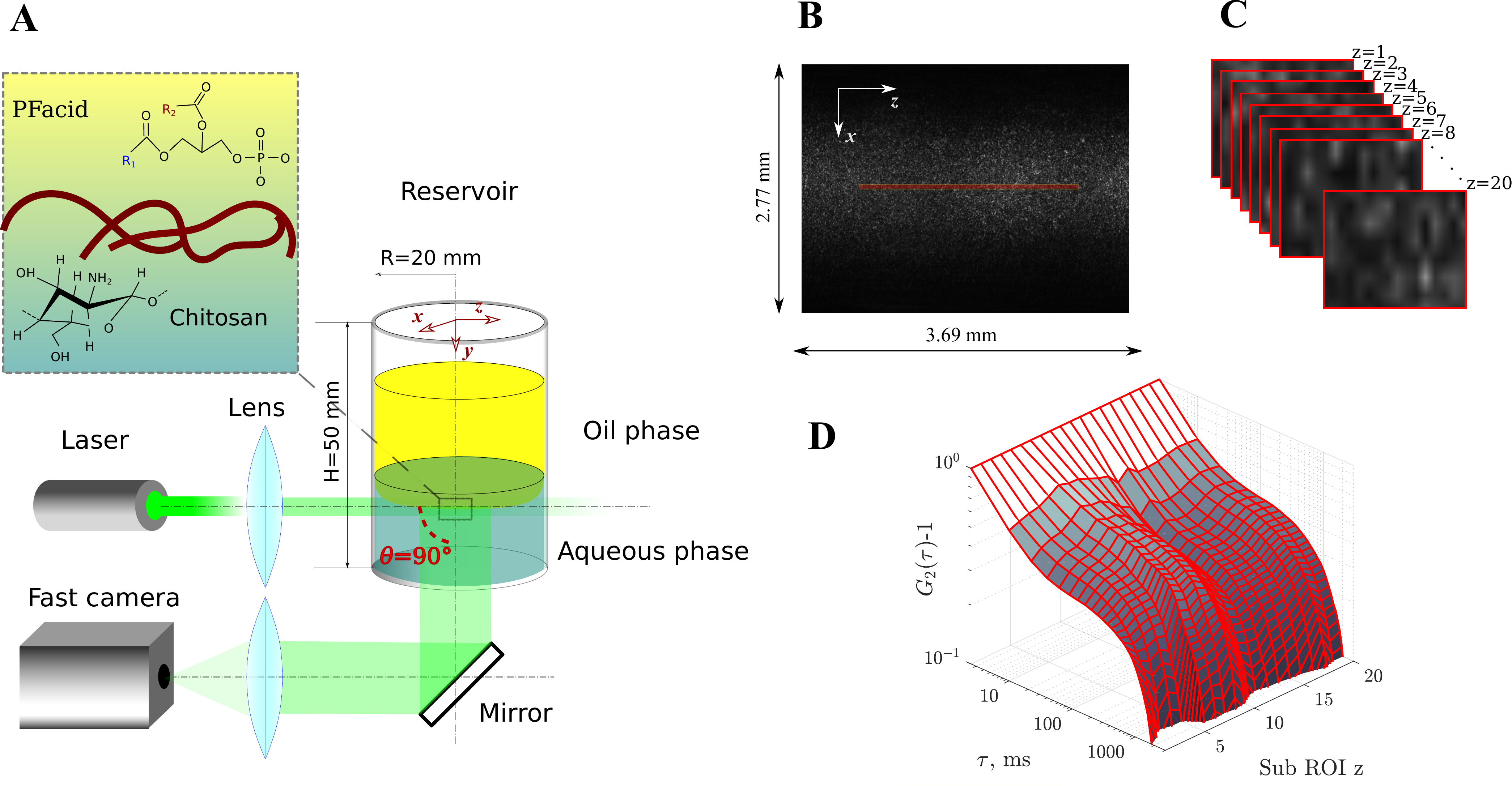}
    \caption{ Representative sketch of multi-speckles Dynamic Light Scattering. (A) Chitosan / PFacid membrane was formed at oil-water interface in a cylindrical container. The interface was illuminated by a laser beam set to propagate inside the membrane. The light scattered at 90$^{\circ}$ was reflected by a polarization holding mirror and collected with a lens onto the camera sensor. (B) Example of an image taken by the CCD camera. In the center, the laser illumination path is clearly visible, showing the speckles. The red rectangle in the middle shows the part of the image used for auto-correlation calculus. (C) Stack representation of the 20$^{\text{th}}$ sub-ROI obtained by sequencing the red rectangle in B in equisized small images. (D) Intensity correlation function $g_2-1$ as a function of the lag time $\tau$ and the sub-ROI $z$.}
	\label{FIG:dls-setup}
\end{figure}

\section{Materials and methods}
\subsection{Materials}

Chitosan (CH) powder with medium molecular weight and 75-85\% deacetylation was purchased from Sigma-Aldrich ($M_w =$ 190–310 kg/mol, CAS number 9012-76-4, Sigma-Aldrich). The anionic surfactant used to form a  complex with chitosan at water-oil interface was phophatidic fatty acid (PFacid).  It was comprised of a commercially available lecithin known as lecithin YN (Palsgaard 4448, food-grade, E442, Palsgaard). In mass, the phosphatidic acids were 55\% w/w, neutral triglycerides 40\% and ammonium salts 5\%. More than 90\% of the fatty acid chains are C18, see \cite{gunes2011tuneable} for details. The molecular structures of both compounds are given in Figure \ref{FIG:dls-setup}-A. Sodium hydroxide (1 mol/L) was purchased from VWR. The oil-soluble fluorescent dye, Hostasol Yellow 3G (HY-3G), was acquired from Clariant. Rapeseed oil (from \textit{Brassica rapa}, CAS number 8002-13-9), hydrochloric acid (36.5-38.0 \%, BioReagent, for molecular biology) and  cyclohexane (anhydrous, 99.5\%, CAS number
110-82-7)  were obtained from Sigma-Aldrich. Deinonized water (resistivity > 18 M$\Omega$.cm) was produced from a Millipore Filter water system.  CellMask™ Deep Red Plasma membrane stain  was obtained from ThermoFisher. All chemicals and solvents used in this study were commercially available and used as received unless stated otherwise. 

The aqueous solution was obtained by dissolving chitosan powder in Millipore water and carefully adjusting  the pH with hydrochloric acid (1\,mol/L) at 3.0 to obtain a solution of 0.1\,\% w/w. The chitosan solution was then filtered to remove undissolved particles through Minisart\textregistered Surfactant-free Cellulose Acetate (SFCA) syringe Filter (pore size 5.0\,$\mu$m). The viscosity of the 0.1\,\% chitosan solution was 8\,mPa$\cdot$s. At pH=3.0, we can considered that all the amino groups are protonated, which corresponds to  $\simeq 10^{3}$ positive charges for one mole of the medium molecular weight chitosan used in this study  \cite{babak2000formation}.

The  1\,\% w/w  stock solution of PFacid  was obtained by  dissolving lecithin YN overnight in rapeseed oil (carefully stirred at 35\textdegree C ). Undissolved particles were removed by centrifugation at 1000 g for one hour. The solution was then diluted with rapseed oil to obtain a concentration of PFacid ranging from 0.1 to 1\,\% w/w. The viscosity of these solutions was 62.6\,mPa$\cdot$s at 23 $\pm$ 1\textdegree C. If not stated otherwise, we used 0.1 \% w/w chitosan solution and 0.1 \% w/w PFacid solution in order to catch assembly phenomena at short time scales. We estimated that this 1:1 ratio of mass corresponds approximately to a 1:1 ratio of free charges. 

\subsection{Interfacial rheometry} 
An interfacial rheological study of a flat film of chitosan / PFacid complex was performed with a bicone geometry using a commercially available solution (\textbf{Figure \ref{FIG:interfacial-rheometry}}), which is an appropriate approach for interfaces with
high moduli and viscosities \cite{renggli2020operating}. The Interfacial Rheology System (IRS , Anton Paar, Austria) was mounted on the Modular Compact Rheometer MCR 501 (Anton Paar, Austria) after being thoroughly washed with ethanol and Milli-Q water. For the interfacial measurements, the bicone geometry was positioned at the height $H_1=$ 19.5\,cm from the bottom of the measuring cell after the zero-gap was established. Then the cell was filled with the aqueous phase until the normal force acting on the geometry was not adjusted to zero point in order to position  the edge of the bicone geometry exactly at the interface.  Next, the oil phase was gently added over the aqueous phase  up to the total height $H =$ 40\,cm. Every measurement was performed in 3-4 minutes after two phases were brought into contact. All oscillatory measurements were performed for at least five time periods per data point. All the measurements were conducted at room temperature (23 $\pm$ 1\textdegree C).

The interfacial viscoelastic properties of the chitosan / PFacid membrane in oscillatory motion are described by  the frequency-dependant complex linear viscoelastic modulus $G_i^*$, 

\begin{equation}
    G_i^*(\omega) = G_i^{'}(\omega) + iG_i^{''}(\omega)
\end{equation}

where $G_i'$ and $G_i''$ are the components of the interfacial complex modulus (two-dimensional elastic modulus and loss modulus, respectively). It is related to the  the complex interfacial viscosity $\eta _i^*$ as \citep{renggli2020operating} 

\begin{equation}
    G_i^*(\omega) = i\omega\eta _i^*(\omega) = -\omega\eta _i^{''}(\omega) +i\omega\eta _i^{'}(\omega) 
\end{equation}

where $\eta _i^{''}(\omega)$  is  the out-of-phase shear viscosity and $\eta _i^{'}(\omega)$  is the  the dynamic interfacial shear viscosity. The contributions of the interfacial and bulk components to the torque appearing on the bicone geometry during its motion were compared through the non-dimensional parameter,   the Boussinesq number ($Bo$)

\begin{equation}
    Bo(\omega) = \frac{\eta _i^{'}(\omega)-i\eta _i^{''}(\omega)}{a(\eta_b^{(1)} +\eta_b^{(2)})}
\end{equation}

where $\eta _b$ is the bulk viscosity (superscripts denote upper and lower fluid respectively) and $a$ is the characteristic length scale that depends on the measuring system. As usual, the interfacial flow was considered to be  decoupled from the bulk. In that case,  the interfacial shear viscosity is  calculated by \citep{bicone_theory_2003}:

\begin{equation}
   \eta = \frac{M-\frac{8}{3}R_2^{3} (\eta_b^{(1)} +\eta_b^{(2)}) \Omega}{4\pi R_2^{2} \Omega}
\end{equation}

where $\Omega$ is the angular velocity (Figure \ref{FIG:interfacial-rheometry} A). This expression is only relevant for the $ Bo\to\infty $. For low and intermediate $Bo$ a complete analysis must be used, since the influence of the bulk phases becomes important \citep{bicone_theory_2003}. 
In our experiments the  interfacial response was decoupled from the bulk one by using the Anton Paar application software. 

\subsection{Particle tracking velocimetry}
The displacements and  velocity field on the oil-water interface during rheometric experiments were quantified through  particle tracking velocimetry (PTV). For this purpose, the water-oil interface was decorated at low coverage with polyethylene microspheres (63-75 µm Cospheric LLC, USA) used as tracers. Less than 0.01\% w/w of particle powder was added to 100 mL of oil phase and mixed thoroughly with a magnetic stirrer overnight. This volume of oil containing tracers was further used for rheological experiments as described above in the Section 2.2. The USB microscope (A1 USB Digital Microscope, Andonstar) was immersed in the oil phase during rheological experiments in IRS in order to visualize the displacement of microspheres under the shear flow. The image sequences were recorded at 20 frames per second and post-processed with a custom written particle tracking routine (MATLAB, MathWorks).

\subsection{Dynamic light scattering}

The dynamic evolution of the structure of the membrane was measured by space- and time resolved DLS at constant temperature T=22\textdegree C. A sketch of the custom-built DLS set-up is shown in Figure \ref{FIG:dls-setup}. The oil-water interface, which later became a membrane, was illuminated by  a vertically polarized laser beam  produced by a single-mode laser (MSLIII, CNI, China, $\lambda =$ 532\,nm). The laser beam had a diameter of 2 mm and was shaped by a combination of two lenses with focal lengths  $f_{1}=$200\,mm and $ f_{2}=$-25.4\,mm. The coherent light was scattered by forming solid matter at the oil-water interface.  Only the light scattered at 90\textdegree  was collected, after reflection onto a non-polarized mirror. Focusing the laser beam on the interface was a complicated technical task, as the oil-water interface formed a concave-convex meniscus  depending on the wettability of the cylinder. However the chitosan/PFacid complexation leading to the membrane formation resulted in a drastic decline in interfacial tension causing the interface to flatten. This led to the interface displacement along \textit{y}-axis and consequently signal loss. In order to minimise this effect, all measurements were performed using a large custom-made glass cylindrical container positioned vertically and sealed underneath with a flat sheet of glass. The dimensions of the reservoir rendered the interface displacement negligible and the precise control of the sample volume ensured the tangential contact between the interface and the laser beam throughout the experiments.  

In order to follow the structural evolution of the membrane, the scattered intensity was collected either with a CCD camera (acA640-100gm, Basler, Germany) or with a photomultiplier (SPCM-AQR-13, excelitas Technologies, USA).  When the camera was used, a lens with a focal length  $ f_{l} =$ 150\,mm allowed the image of the scattering volume to form onto the CCD sensor. A diaphragm placed in the focal plane of the lens was set in order to optimize the size of the speckles to the pixel size of the camera \citep{Ali2016PhD}. For fast processes, the photomultiplier associated with a correlator (Flex03-LQ, Correlator.com, USA) was used to widen  the dynamic range of acquisition to include lag times as small as 10$^{-6}$\,s. The wave vector $q$ is defined as $4 \pi / \lambda \sin(\theta / 2)$ with $\lambda$ the wavelength of the laser in the scattering medium and $\theta$ the angle of observation. As we were probing the oil-water interface, we considered the averaged optical index of both phases to calculate $\lambda$ (1.333 for water and 1.471 for the oil).  Consequently, the wave vector $q$ was 23 $\pm$ 2 $\mu$m$^{-1}$.
 
Our approach enabled nondestructive probing of the interfacial membrane evolution with both spacial and temporal resolution, as long as the characteristic relaxation time of the studied system allows signal detection with a digital camera. The scattered light detected by CCD camera created the image of a coherence area known as speckle (Figure \ref{FIG:dls-setup} B).  The red rectangle in the center of the image  represents the Region Of Interest (ROI), only this part of the image has been used for analysis. This area was sequenced into 20 sub-ROI (Figure \ref{FIG:dls-setup} C). The laser beam passing through the interface created a trace that was visualised by CCD camera as bright area in the middle of the image along z-direction (see Figure 2 B). The region of interest was chosen in the middle of that area and was limited by 10 pixels along x-direction. This choice was justified by a compromise between the image size and the readout speed of CCD camera. The individual  time autocorrelation function of the scattered intensity $g_2(\tau) -1$ was computed for each sub-ROI

\begin{equation}
   g_2(\tau,z) -1 = \frac{\langle I^{z}(t)I^{z}(t+\tau) \rangle_t}{\langle I^{z}(t)^{2} \rangle_t} -1
\end{equation}

where $I^{z}(t)$ is the intensity collected within $z^{th}$ sub-ROI and $\langle ... \rangle_t$ denotes averaging over time. 
Figure \ref{FIG:dls-setup} D shows the result of the intensity correlation function as a function of the sub-ROI $z$ and the lag time $\tau$. 

We also used the Time Resolved Correlation scheme (TRC) which allows DLS investigation of heterogeneous dynamics, as introduced by \citep{2002Cipelletti, bissig2003intermittent, 2005Duri}. Analogously to $g_2(\tau) -1$, the correlation degree   $c_I(t, \tau, z)$ was calculated individually for each sub-ROI $z$

\begin{equation}
   c_I(t, \tau, z) = \frac{\langle I^{z}_p(t)I^{z}_p(t+\tau) \rangle_p}{\langle I^{z}_p(t) \rangle_p \langle I^{z}_p(t+\tau) \rangle_p} -1
\end{equation}

where $p$ is a pixel of the sub-ROI $z$, $I^{z}_p$ the pixel intensity  and $\langle ... \rangle_p$ denotes averaging over pixels in the sub-ROI.

\subsection{Microscopy}

SEM was used to characterize the morphology of chitosan / PFacid membrane. The membranes were grown on the surface of chitosan drops suspended in oil phase which contained PFacid. Once the required complexation time was achieved the droplets were washed with a large quantity of cyclohexane  (for more details see \cite{gunes2011tuneable}, \cite{xie2017interfacial}) in order to remove the oil with the residues of anionic surfactant. The chitosan droplets encapsulated with the membrane were placed on a cover slip and dried at room temperature.  Dried  chitosan / PFacid membrane were observed with SEM. Samples were coated with Au/Pd in a Baltec MED-020 sputter coater and observed in secondary electron mode in a Thermo Scientific Quanta 250 microscope equipped with a field emission gun and operating at 2.5 kV.

CSM was also used to characterize the morphology of chitosan / PFacid membrane in wet conditions. Analogously to the SEM characterization described above, the chitosan droplets were injected into oil phase containing anionic surfactant. Wet (no cyclohexane washing) chitosan / PFacid membrane were observed with Leica TCS SP8 scanning point confocal microscope equipped with a $\times$63 oil immersion objective and in-plane image resolution 0.36 µm/px.

\begin{figure}
	\centering
    \includegraphics[scale=.5]{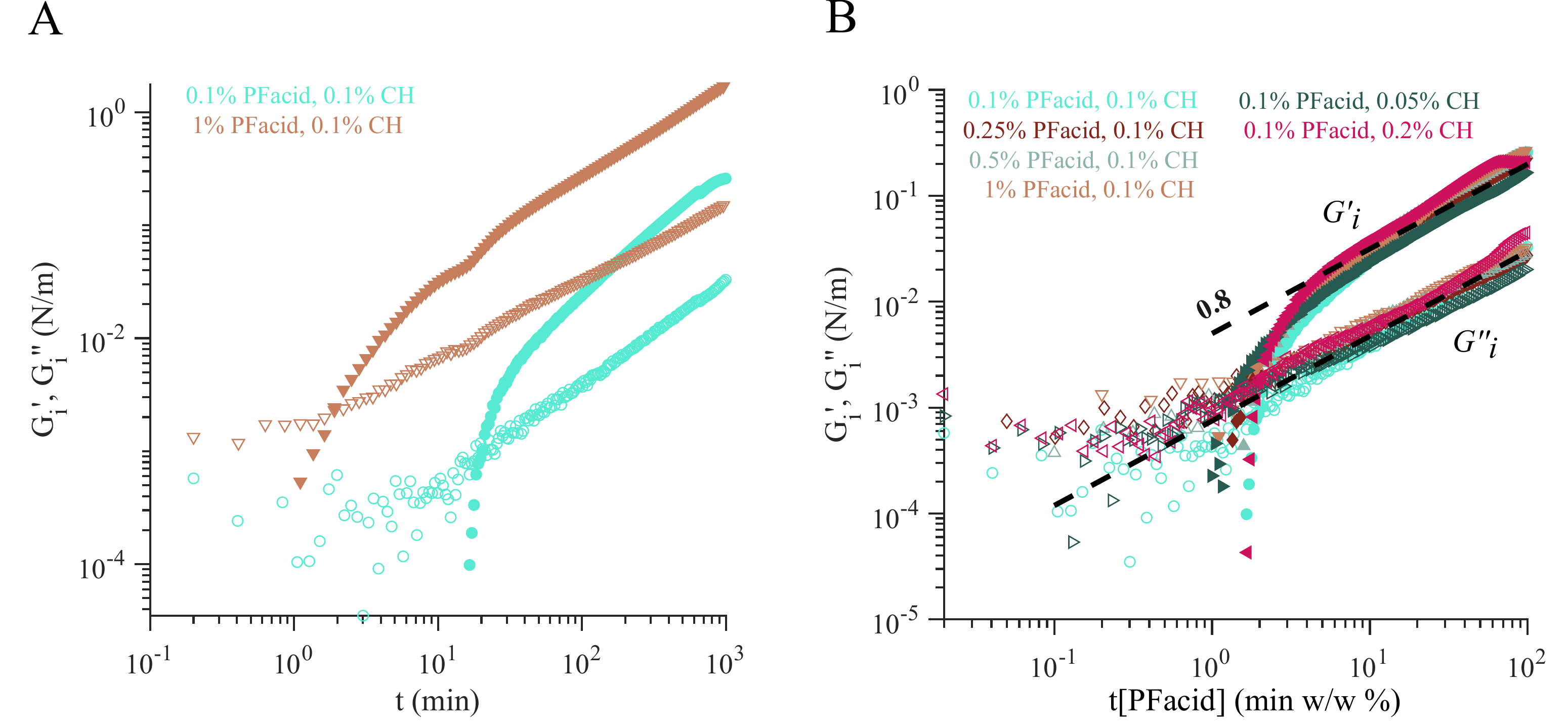}	\caption{  Macroscopic study of the  interfacial rheological properties of a chitosan / PFacid membrane. (A)  Typical time-dependant evolution of interfacial elastic $G'_i$ (filled symbols) and viscous $G''_i$ (empty symbols) shear moduli at different concentrations of PFacid. $f=$ 0.5 Hz, $\dot{\gamma}=$ 0.03\%, chitosan 0.1\% w/w  (B)  The kinetics of $G'_i$ and $G''_i$ collapsed on the same master curves when the time axis was multiplied by the concentration of PFacid, for various concentrations of PFacid and chitosan.  }
	\label{FIG:TS}
\end{figure}

\section{Results and discussion}
\subsection{ Interfacial rheology of chitosan / PFacid membrane}

We analysed the kinetics of formation of the membrane with a time sweep experiment at  constant amplitude ($ \dot{\gamma} $ = 0.03\%) and frequency ($f$ = 0.5\,Hz). The choice of these parameters was justified  by a compromise between signal sensibility and minimizing the disturbance of the interfacial complexation process by the stress. Moreover,  this set of parameters allowed us to keep the deformation within the linear viscoelastic regime at long time scales (after 16h, see Figure SI 1). However, as explained below, the growth of the membrane was disturbed by the applied strain. Figure\ref{FIG:TS}-A depicts the evolution of $G'_i$ and $G''_i$ over time for two different concentrations of PFacid. Note that the 1:1 mass ratio corresponds  approximately to a 1:1 ratio of free charges. As a control, a pure water-oil interface without membrane formation was also quantified (see Figure SI 2), which showed a constant $G''_i$ of  $\sim$  10$^{-3}$\,N/m whereas $G_i'$ was null. In the early stage of membrane formation ($t<$ 1 min for 1\% w/w PFacid and 10 min for 0.1\% w/w PFacid),  $G_i''$ was almost constant and close to the system without PFacid. $G_i'$ was out of the measurement sensitivity. In this regime, the interface manifested purely liquid-like properties. However, within a few minutes, a slow increment in $G_i''$ was accompanied by a rapid growth of $G_i'$. The interfacial storage modulus $G_i'$ quickly overcame  $G_i''$, manifesting the prevalence of solid-like properties. On long time scales, both interfacial modulii increased.

\begin{figure}
	\centering
    \includegraphics[scale=.3]{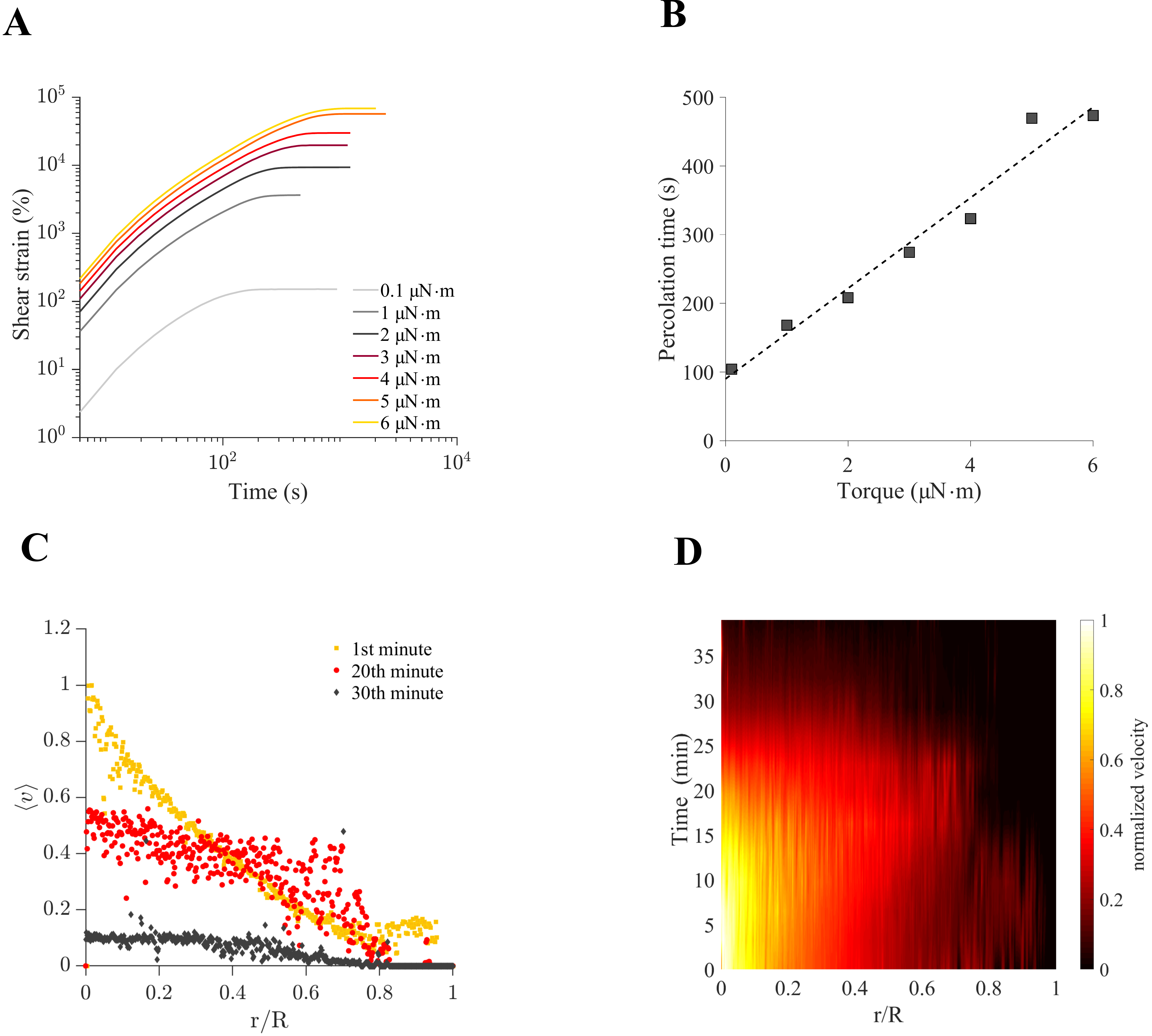}
	\caption{Particle tracking on the water-oil interface during the creep experiment.Chitosan 0.1\% w/w , PFacid 0.1\% w/w.} (A) Creep experiments at the water-oil interface during the membrane formation at different torque values. (B)  The time required to stop the rotation of the geometry increased on a roughly linear basis with the torque. (C) Normalized velocity profiles at the interface at different stages of membrane formation. $r/R=$ 0 is the edge of the bicone. (D) The heatmap of normalized velocity values at the  different points in time.  
	\label{FIG:ptv}
\end{figure}

To gain insight into the mechanisms at  play in the early moments of membrane formation, creep experiments on the forming membrane were coupled to  visualisation of the deformation of the interface by PTV (Figure \ref{FIG:ptv}).  In these creep experiments, the bicone geometry was put into motion at fixed torque values and the deformation was measured. As the membrane was forming, the shear strain increased gradually  until the geometry was brought to arrest. The evolution of the deformation varied with the applied torque, Figure \ref{FIG:ptv}-A. In analogy with percolation of particle laden interfaces \citep{thijssen2017interfacial}, we termed the time at which the strain rate was null, the percolation time. The percolation time increased on a roughly linear basis with the applied torque, Figure \ref{FIG:ptv}-B. Thus, the percolation process of the interface was coupled to the interfacial shear rate. We deduced from the intercept of the linear fit at zero torque, that the percolation time in absence of flow disturbance tended to 100 s.

The water-oil interface was decorated at low coverage with polyethylene microspheres ($\sim$ 70\,$\mu$m)  used as tracing particles.  The radial velocity profile $v$ of the interface was parabolic during the first few minutes of reaction, as expected for liquid interfaces.   The spatio-temporal evolution of the velocity shows that the geometry slow-down was associated with the flattening of the velocity profile, Figure \ref{FIG:ptv}-C,D. Macroscopically, we observed that the shear rate tended towards zero in regions closed from the geometry ($r =$ 0 and $R$). However, the velocity distribution was strongly heterogeneous before the arrest of the geometry.  In fact, a closer look at the interface showed a constant formation and rupture of the membrane. We observed millimetric patch-like sheet  membranes that grew all over the interface and accumulated close to the geometry (see \textbf{Supporting video}).  When the amount of membrane pieces was high enough to fully cover the interface, the interface jammed and stopped the motion of the geometry.  This result fitted with non-reactive particle laden-interface for which domains of packed particles create elastic interfaces. When these domains start to break-up, a transition to viscous-like behavior was observed \citep{barman2016role}. 

Then, we repeated the time sweep experiments for various concentrations of PFacid and chitosan. We observed that the kinetics was mainly dependent on the PFacid concentration. In Figure \ref{FIG:TS}-B, we multiplied the time axis by the concentration of PFacid. We observed a good collapse of the different curves for both interfacial moduli. We deduced from this master-curve that the percolation time increased linearly with the concentration of anionic surfactant. On long-time scales, both moduli scaled with $(t$[PFacid]$)^{0.8}$ (dashed lines), whereas the chitosan/PFacid film thickness scales with the square-root of the time (see \textbf{Supporting material}). We could interpret this as a non-homogeneous growth of the membrane. We have to be cautious with the interpretation of the slope of viscoelastic moduli, as it was not possible to carry out the measurements within the linear viscoelastic regime during the course of membrane formation. However, all these features indicated that the growth of the membrane at long time scales was limited by the diffusion of the surfactant in the membrane under formation, similarly to interfacial polymer complexation \cite{de2020entropically}. 

Finally, we concluded that the complexation of chitosan with PFacid is a two-step process. At short time scale, the interface has a macroscopic rheological behaviour which is characteristic of liquid interfaces, but the interface is strongly heterogeneous and is composed of solid patches. At a critical time, termed the percolation time, the millimetric patches pave the interface and  percolate, which is  characterized by a sudden increase of the interfacial elastic modulus $G'_i$. This process depends on both the hydrodynamic conditions and the concentration of the diffusing surfactant. These observations  are also  analogous to  gelation of polymer under shear, for which there is a competition between the formation of clusters that tend to form a percolated network and hydrodynamic forces that disrupted the network \cite{deCarvalho1997physical}. At long time scales, the thickness of the membrane grew by diffusion of the surfactant in the membrane under formation, as observed during the interfacial complexation of polymers \cite{de2020entropically}.

\subsection{Dynamic light scattering experiments}

\begin{figure}
    \centering
    \includegraphics[width=0.5\textwidth]{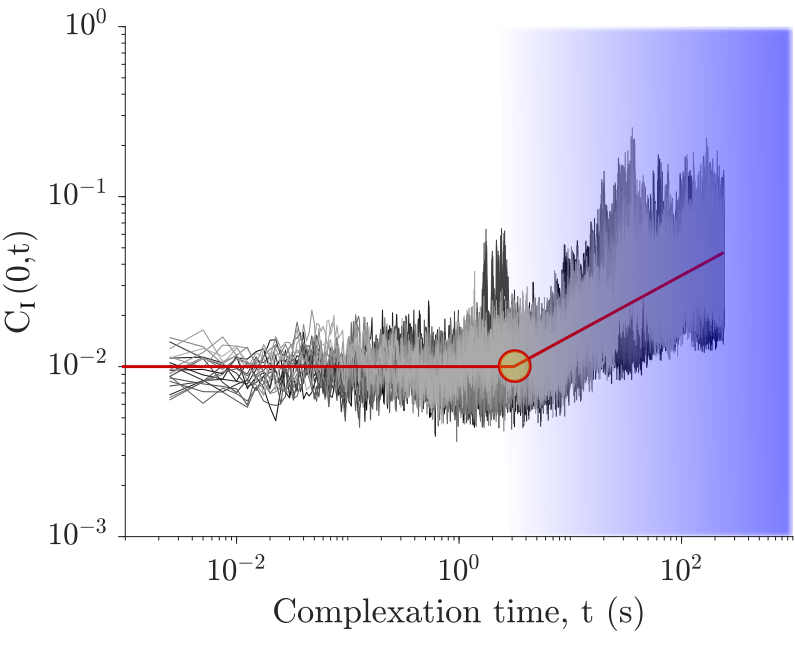}
\caption{Time Resolved Correlation (TRC) at 0 lag time ($\tau$) of building interface of the 20 ROIs. Straights lines and the color background are eye-guides and the circle at line interception indicates the starting time of the membrane gelation.Chitosan 0.1\% w/w, PFacid 0.1\% w/w.}
\label{FIG: Ci}
\end{figure}

In order to shed light on the structural evolution of the interface formation at a smaller length scale, we employed DLS and time-resolved correlation (TRC) analysis \cite{2002Cipelletti, bissig2003intermittent, secchi2013biopolymer}, see Figure \ref{FIG:dls-setup} for the experimental set-up. The length scale that we probed with our set-up was of 40-45\,nm.The interface was formed at the water-oil interface in a cylindrical reservoir via complexion between 0.1\,\% w/w chitosan and 0.1\,\% w/w PFacid, Figure \ref{FIG:dls-setup}. During the first 2\,s, $C_I(\tau=0,t)$ fluctuated randomly around a steady value of $\simeq 10^{-2}$, Figure \ref{FIG: Ci}. Such behaviour corresponds to a Brownian system \citep{2002Cipelletti,2005Duri}, meaning that the  displacement of particles between any two frames was on average the same, independently of the complexation time $t$. Beyond 2\,s, the resolved correlation function drastically increased and gained half a decade in 100\,s, meaning that the degree of correlation in the sample increased. This behaviour indicated the formation of a gel-like interface at the scale of 40-45\,nm \cite{2002Cipelletti}.   

\begin{figure}
    \centering
    \includegraphics[scale=.35]{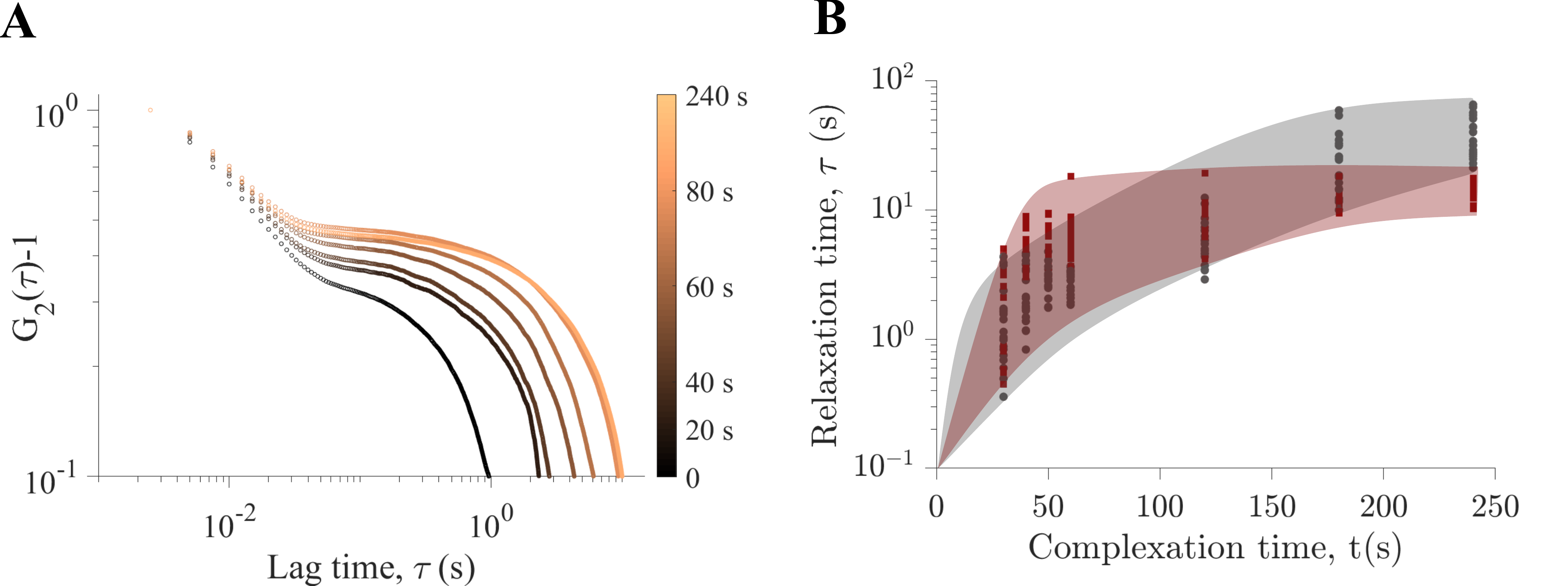}
    \caption{(A) Auto-correlation function of the interface computed from the mean time of TRC as at different lag times for an ROI as a function of the time from 0 to 240\,s. (B) Evolution of the relaxation times of the forming interface within different sub-ROI computed for two separate  experiments at identical conditions. The colored areas separating two experiments serve as guiding lines. Chitosan 0.1\,\% w/w, PFacid 0.1\,\% w/w.}

\label{FIG: G2}
\end{figure}

The kinetics of membrane formation was also probed with the intensity correlation function $G_2(\tau)-1$. Figure \ref{FIG: G2}-A shows the intensity correlation function computed at different lag times following the  formation time of the membrane from 0 to 240\,s. At 0\,s, the intensity correlation function was identical to the correlation of the chitosan alone (see \textbf{Supporting material}) with a relaxation time of 0.1\,s. As the time increased, a second relaxation time appeared as a second mono-exponential function for which relaxation time increased from 0.1\,s to more than 10\,s in 200\,s of membrane complexation. After long complexation (10 hours),   the characteristic relaxation time of a membrane increased up to 2$\times$10$^{3}$ (see \textbf{Supporting material}). 

The results of spatial analysis are depicted in  Figure \ref{FIG: G2}-B, where each dot corresponds to an ROI of 56\,µm x 56\,µm. Initially,  the relaxation time at the interface was the same as that of the chitosan. As the chitosan / PFacid complexation took place and a solid matter started to appear at the interface, the relaxation time increased. At the different complexation times considered here, the relaxation times differed by nearly one decade between different ROI. It indicated high dynamic heterogeneity in the interface complexation. This was consistent with the observation of patches during the interfacial rheological measurement. We concluded that interfacial complexation of chitosan with the anionic surfactant PFacid is a spatially heterogeneous process.

\begin{figure}[tb!]
	\centering
    \includegraphics[width=1\textwidth]{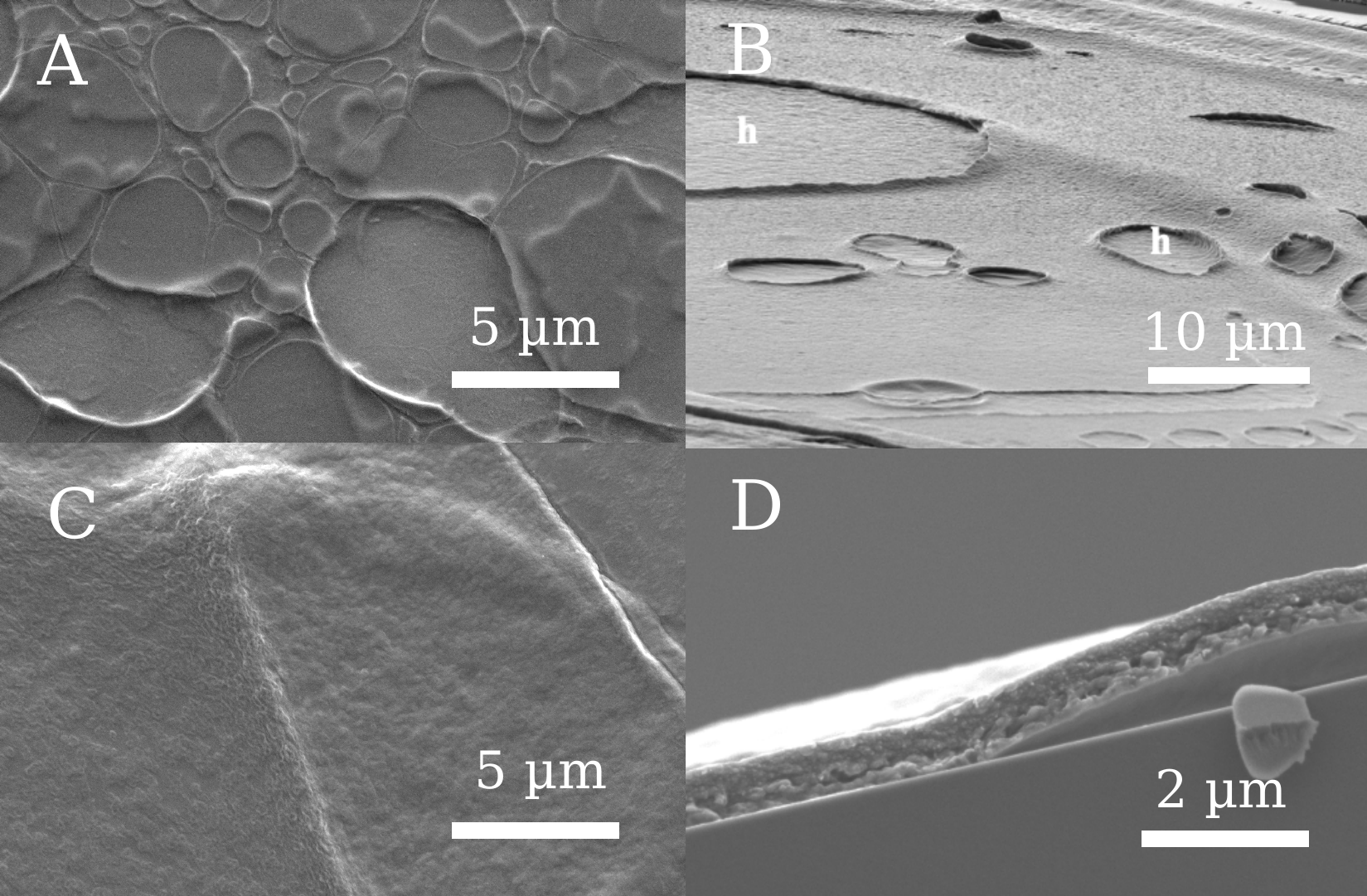}
	\caption{SEM images of dried chitosan / PFacid membrane.  (A)  CH / PFacid membrane after 30\,s of complexation. The interface was made of individual patches of 1-50 $\mu$m size (p), as well as large holes (h). (B) CH / PFacid membrane after 2 min of complexation. The interface was homogeneous, with the exception of circular holes (h).
	(C) CH / PFacid membrane after 2 min of complexation. The interface is fully formed though characterized by a certain roughness .(D) View in the thickness of the membrane after 2 min of complexation. The membrane showed a granular structure.Chitosan 0.1\,\% w/w, PFacid 0.1\,\% w/w.}
	\label{FIG:morphology}
\end{figure}

\begin{figure}[tb!]
	\centering
    \includegraphics[width=1\textwidth]{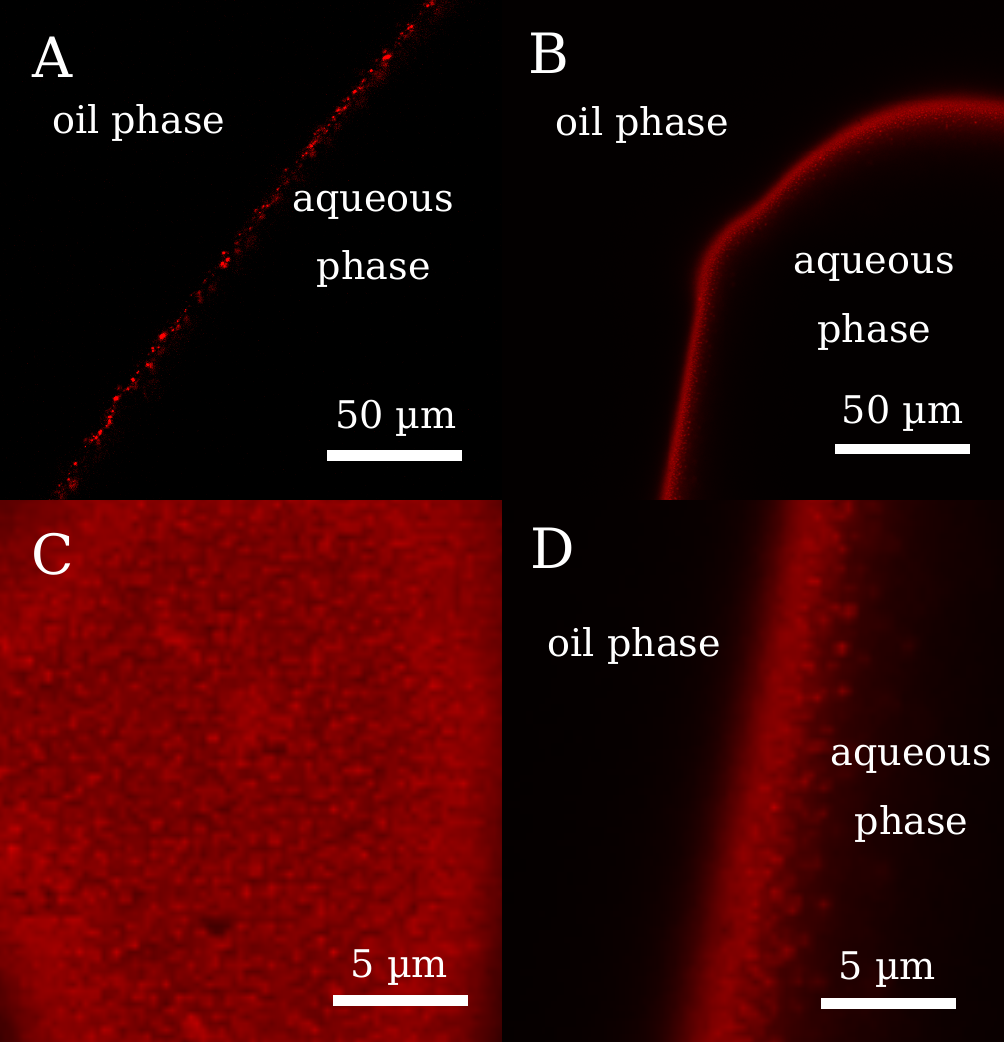}
	\caption{Confocal images of  wet Chitosan / PFacid membrane, the membrane was marked by a fluorescent dye with high lipid affinity (see text for details).  (A) lateral optical slice of a membrane after 1 minute of complexation. (B) lateral optical slice of a membrane after  48h of complexation.  (C) A piece of a membrane laying flat on a glass substrate showing a granularly patterned structure, 48h of complexation (D) A horizontal confocal slice of a labeled membrane, 48h of complexation. Chitosan 0.1\,\% w/w, PFacid 0.1\,\% w/w.}
	\label{FIG:confocal}
\end{figure}

\subsection{Scanning electron and confocal microscopy}

The morphology of chitosan / PFacid membranes was observed with SEM. In order to minimise harsh manipulations with fragile membranes, chitosan / PFacid membranes were grown on a surface of water  droplets in oil phase containing chitosan and PFacid respectively, for more details see \cite{xie2017interfacial}. The complexation reaction was stopped by a gentle washing in large quantities of cyclohexane. After that, the droplets now enclosed by a solid membrane were placed on the glass substrate and dried prior the SEM imaging.

Figure \ref{FIG:morphology} demonstrates the morphology of the chitosan / PFacid formed at 0.5 and 2\,min.  The membrane formed at 0.5 min (Figure  \ref{FIG:morphology} A) was characterized by an important heterogeneity of its structure. It appeared to be formed out of a large number of non-connected patches of 1-5\,$\mu$m. Additionally, large non-circular holes up to 10\,µm were found in the membrane. Although these large perforations of the membrane could be attributed to the damage during sample preparations, the presence of non-connected pacthes fitted with our previous analysis. At 2 minutes of complexation the membrane appears to be intact  although the  presence of large circular holes can be noted on occasion (Figure \ref{FIG:morphology} B), that fit with a percolated network.  The circular shape of the voids could be due to the effect of surface tension. Figure  \ref{FIG:morphology} C shows a magnification of the same membrane which was quite granular. Figure  \ref{FIG:morphology} D  shows the transverse view of the membrane, which was characterized by sub-micrometric aggregates.  

The confocal imaging was carried out at 1 minute complexation time in order to visualize the initial stages of membrane formation and on the thick membrane after 48h of complexation in order to reveal the internal structure of the interface.  Water droplets containing 0.1\,\% w/w concentration of chitosan ware injected into the oil phase containing 0.1\,\% w/w concentration of  PFacid. Water-soluble fluorescent dye with high lipid affinity (CellMask™ Deep Red plasma membrane stain, Invitrogen™) was added to the aqueous phase. This dye has little to no fluorescence in a free form and is only fluorescent once it is "anchored" to lipids. Figure \ref{FIG:confocal} depicts the results of confocal imaging.  Note that the environmental conditions of observations were very different from SEM imaging, where the membrane was dried before visualization. After 1 min of complexation, the side view of the interface shows the presence of discrete nano-metric patches, Figure \ref{FIG:confocal}-A. At long time scales, these inclusions formed a continuous membrane,  Figure \ref{FIG:confocal}B-D.  Figure  \ref{FIG:confocal}-D shows the concentration gradient of these inclusions from the oil phase towards the aqueous phase. As the fluorescent dye  anchored to lipids, we deduced that this gradient of light intensity corresponds to a gradient of PFacid.  

This set of microscopy images consolidated the idea that membrane formation was due to the percolation of individual patches. When the patches were able to form a percolated network, large  holes were present. These results were reminiscent of interfaces covered by model nanoparticles which form heterogeneous structures with voids for low particle surface coverage \citep{masschaele2009direct,reynaert2007interfacial,thijssen2017interfacial}. The basic bricks are sub-micrometric aggregates of chitosan/PFacid. On long time scales, the membrane was fully covered of these aggregates. In the thickness of the membrane, there was a negative gradient of these aggregates from the oil phase to the water phase. This last result supports the idea that the growth of the membrane was limited by the diffusion through a gel-like porous network of PFacid on long time scales, as described for  H-bond donor / acceptor polymers \citep{de2020entropically}.

\section{Conclusion}

Membrane formation based on the complexation between chitosan and short chain fatty acid has been used as a model for interfacial self-assembly of polyelectrolytes and charged surfactant. A multi-scale approach was used in order to perform a characterisation of membrane formation and morphology. 

The multiscale approach created a robust overaching picture of the interfacial complexation process that we sum-up here. The basic bricks of these membranes are sub-micrometric aggregates of chitosan and surfactant, that were observed by SEM and CSM (Figure \ref{FIG:morphology}, \ref{FIG:confocal}). The structure of these aggregates is an open question. At the early stage of membrane formation, the interface behaved like a fluid at macroscopic scale (Figure \ref{FIG:interfacial-rheometry}-a), and the spatial degree of correlation was low at length scales of 40-45 nm (Figure \ref{FIG: Ci}-a). Then, solid patches emerged and increased in size and/or number, which was visible at different length scales. At the smallest length scale used in this study, this corresponded to a sudden increased of the spatial correlation probed by DLS  (Figure \ref{FIG: Ci}-a). At micrometric scales, individual patches were visible by SEM (Figure \ref{FIG:morphology}-a). At macroscopic scales, millimetric patches transported by the interfacial flow were also visible in rheometric experiments. The elastic behaviour emerged suddenly when the patches percolated (Figure \ref{FIG:interfacial-rheometry}, \ref{FIG:ptv}). The formation of the percolated newtork was also supported by SEM images that showed the presence of a continuous network with voids (Figure \ref{FIG:morphology}-b). The process of percolation was mainly dependent on the interfacial stress and the concentration of PFacid. On long time scales,  the growth of the membrane was limited by diffusion of the surfactant.

In conclusion,the interfacial structuring of chitosan with the anionic surfactant is strongly analogous to bulk gelation of polymers \cite{deCarvalho1997physical} and shared some common features with different interfacial systems such as nanoparticles, \cite{cui2013stabilizing,masschaele2009direct,thijssen2017interfacial} polymer or polyelectrolytes \citep{de2017one,le2015interplay,de2020entropically}. Finally, this work brings new elements on interfacial complexation that  should prove useful to control the properties of liquid-liquid interfaces for the design of new materials, such as microcapsules or structured liquids. 


\section{Acknowledgements}
LRP is part of the LabEx Tec21 (ANR-11-LABX-0030) and of the PolyNat Carnot Institute (ANR-11-CARN-007-01).  The authors express their sincere gratitude to Christine Lancelon-Pin (CERMAV-CNRS, Grenoble, France) for her assistance with scanning-electron microscopy. The authors acknowledge the NanoBio-ICMG chemistry platform (UAR 2607, Grenoble) for granting access to the electron microscopy facilities. The authors thank the support of ANR 2DVisc (ANR-18-CE06-0008-01).

\section{Author contributions}
\textbf{Revaz Chachanidze:} Conceptualization, Methodology, Formal analysis, Investigation, Writing - Original Draft, Visualization
\textbf{Kailie Xie:} Conceptualization, Methodology,  Investigation, Writing - Review \& Editing 
\textbf{Hanna Massad:} Methodology, Formal analysis, Investigation, Writing - Original Draft, Visualization
\textbf{Denis Roux:} Methodology, Formal analysis, Writing - Original Draft, Supervision
\textbf{Marc Leonetti:}  Conceptualization, Methodology, Writing - Original Draft, Supervision
\textbf{Clément de Loubens:} Conceptualization, Methodology, Writing - Original Draft, Supervision

\bibliographystyle{elsarticle-num}
\bibliography{main}

\end{document}


\begin{center}
    
\large{Structural characterization of the interfacial self-assembly of 
chitosan with oppositely charged surfactant} 

Supplementary informations

by R Chachanidze \emph{et al.}
\end{center}

\section{Interfacial rheometry}

\begin{figure}[!h]
	\centering
    \includegraphics[width=1\linewidth]{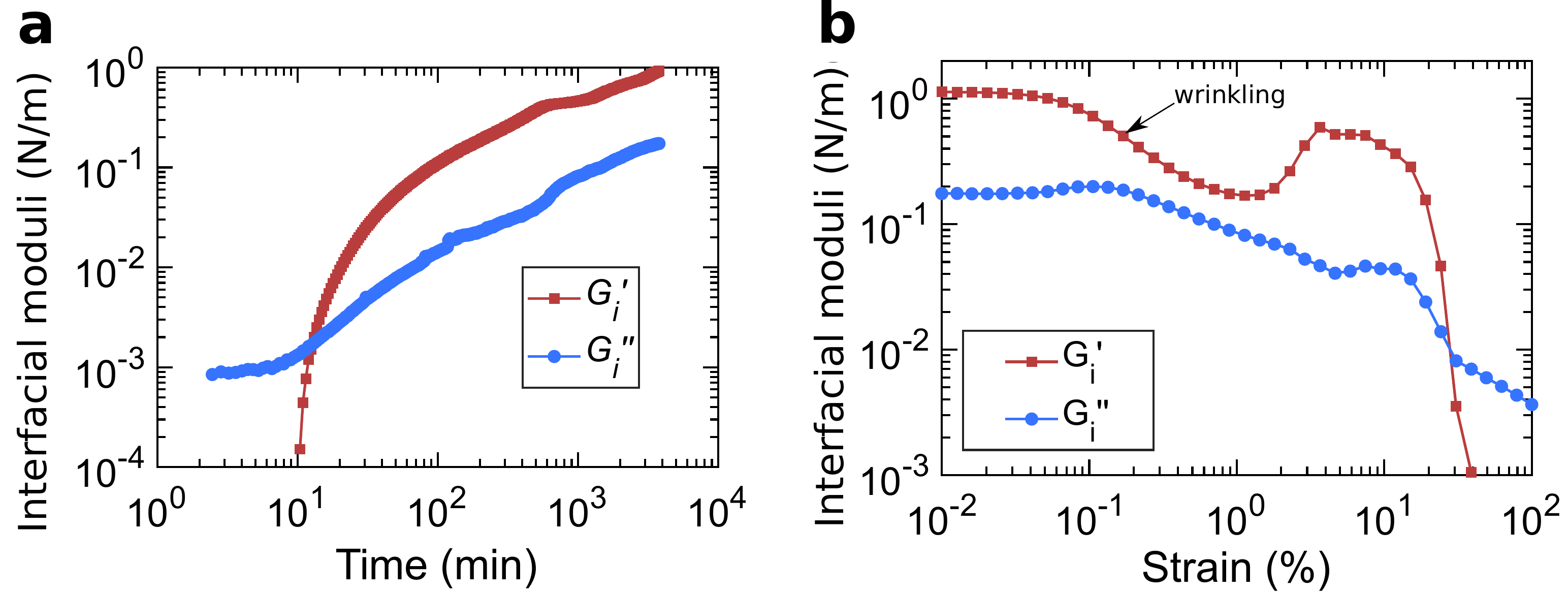}
	\caption{\textbf{a:} We performed a long term membrane formation for a single case (0.1\% w/w chitosan,   0.1\% w/w PFacid , $f$= 0.5 Hz, $\gamma$=0.03\%) in order to perform a strain sweep on a thick membrane and define the limits of linear  elastic regime. \textbf{B:} We observed that for long term formed membranes (>24 hours) the linear elastic regime was limited to surprisingly low values of strain (<0.1 \% ). The direct microscopic observation of the membrane decorated with tracing particles during the strain sweep experiments showed that this highly non-linear behaviour was caused by wrinkling instability of the membrane. The deformation of $\approx$ 20 \% lead to the rapid decline in $G_i'$. The direct observation showed that it was caused by the loss of connectivity between the membrane and the bicone.}
\end{figure}

\begin{figure}[!h]
	\centering
    \includegraphics[width=.8\linewidth]{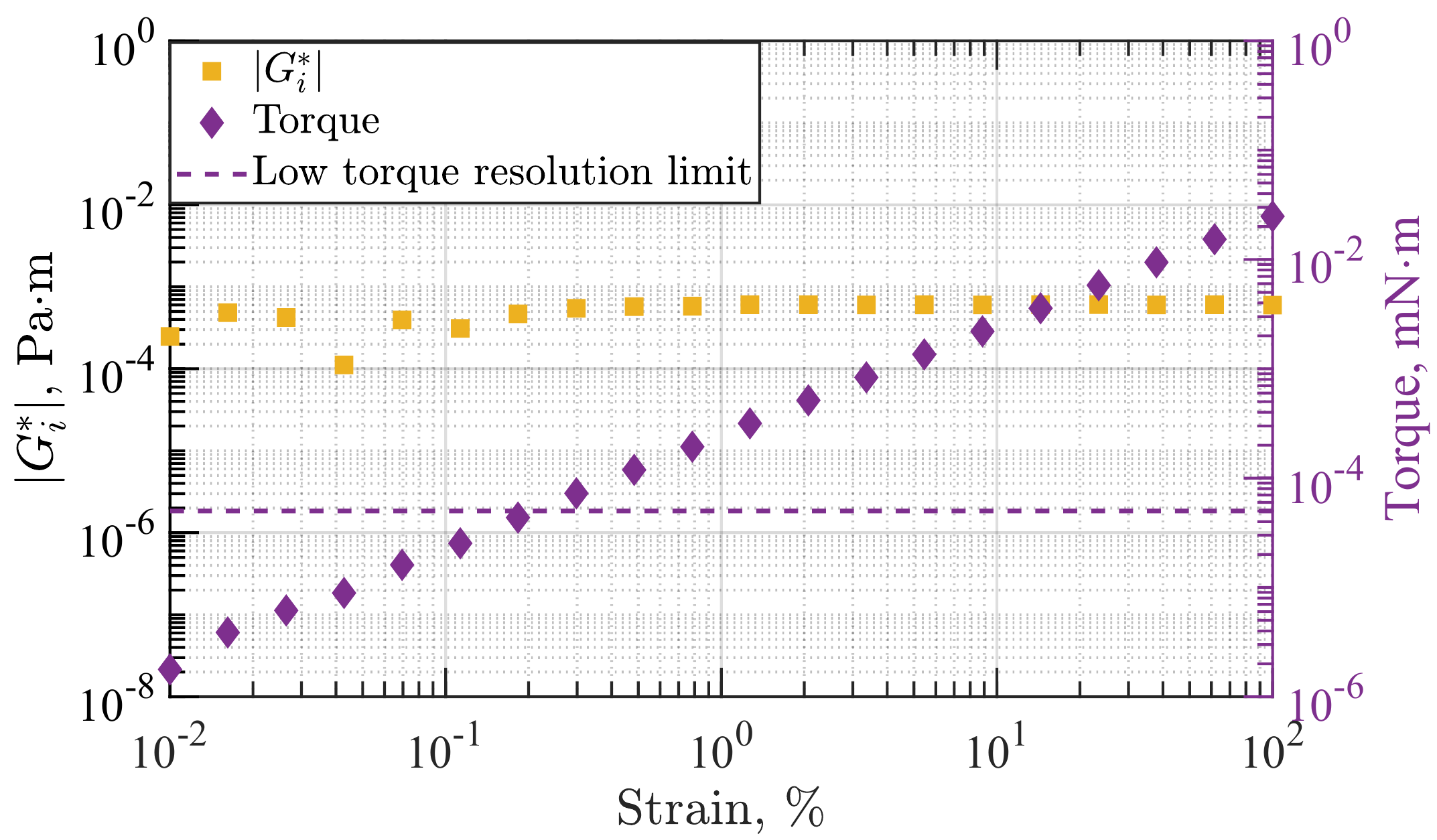}
	\caption{Strain sweep of the chitosan/oil interface. The test was performed with the bicone geometry ($f=$0.5 Hz). The aqueous phase contained 0.1\% w/w chitosan. PFacid was not added to the oil phase. The result illustrates the steady state chitosan/oil interface without the complexation. \textbf{Note}: while the stable values of interfacial viscoelastic modulus $G^*_i$ are obtained, without the solid membrane forming at the interface the Boussinesq number is very low ($Bo$=0.242 in this case). Thus this result is qualitative and served as a  reference line.  }
\end{figure}

\clearpage

\section{Dynamic light scattering}
\begin{figure}[!h]
	\centering
    \includegraphics[scale=.7]{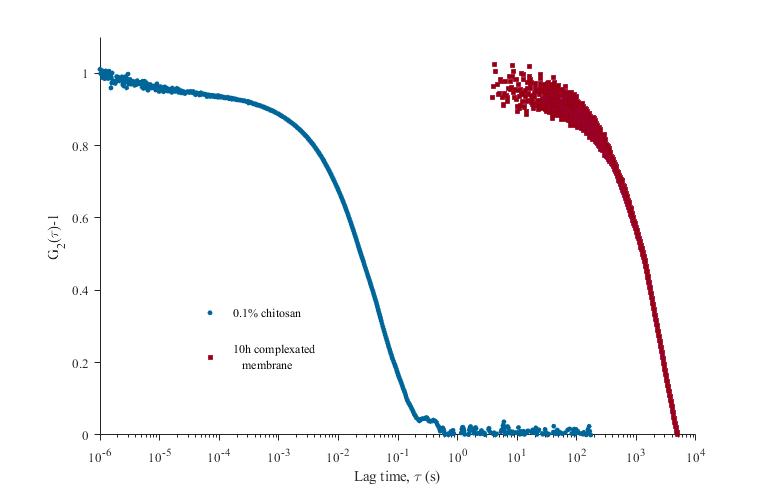}
	\caption{Two distinct cases demonstrating the changes in chitosan/PFacid membrane DLS signature. The average characteristic relaxation time of chitosan solution (0.1\% w/w) was $\tau =$ 0.88 s. The relaxation time of fully formed membrane after 10 hours of complexation was 3488.4 s. Note, that DLS signature of the chitosan solution, being a fast process, was acquired using the PM as a receiver, while the DLS signature of the membrane was acquired with fast camera.}
\end{figure}

\clearpage

\section{Membrane thickness measurement}

\begin{figure}[!h]
	\centering
    \includegraphics[scale=.6]{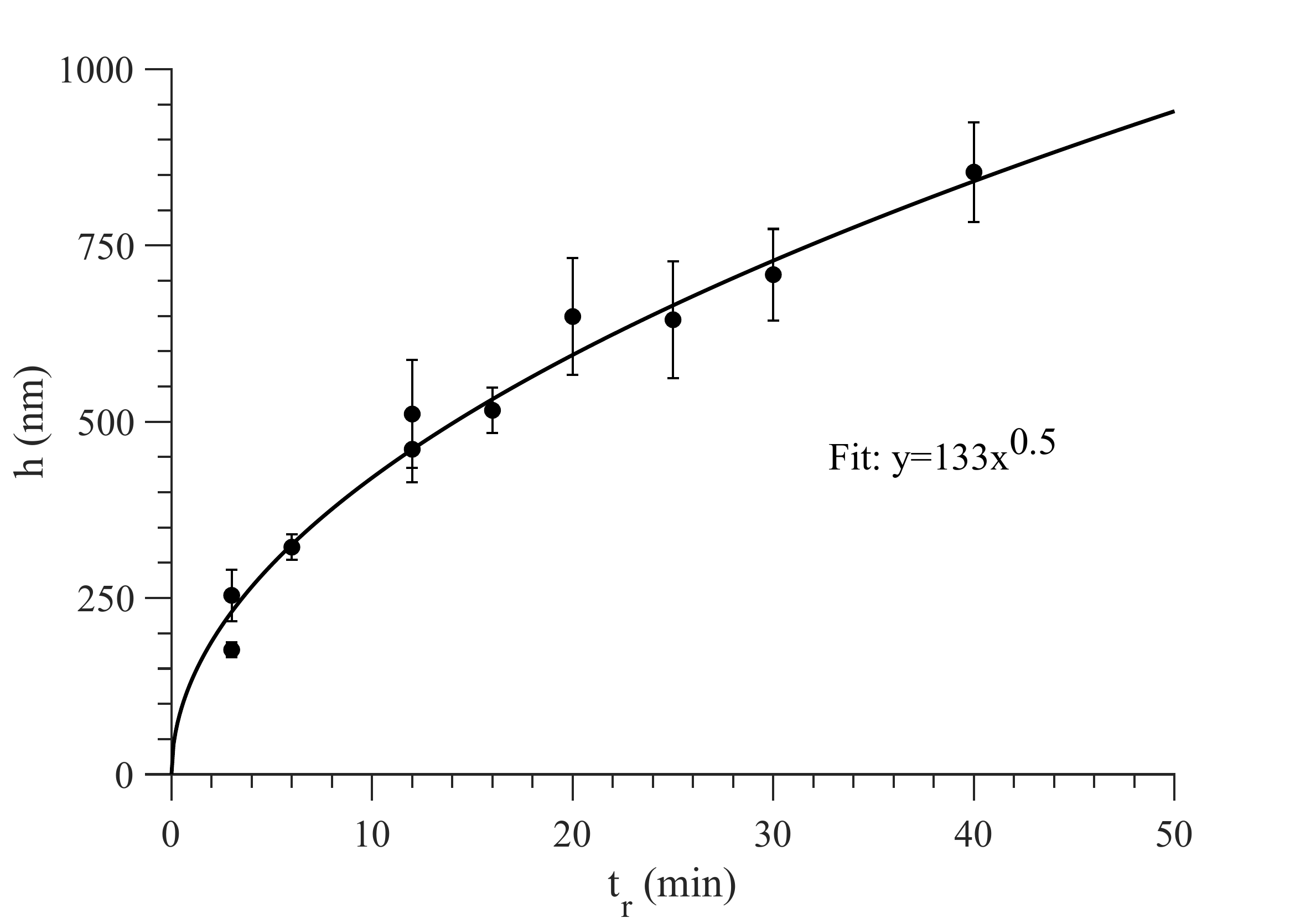}
	\caption{Growth of chitosan/PFacid membrane thickness as a function of reaction time. The square root fitting is consistent with the diffusion driven process. The membranes were grown on a surface of water droplets in oil phase containing chitosan and PFacid respectively as described in the section. The membrane thickness is estimated by measuring the height of the collapsed flat regions on the encapsulated droplets. For SEM, the  capsules were dried on a glass with a conductive coating (ITOSOL12 from Solems France). SEM imaging are carried out to image the cross section of membrane using a GeminiSEM 500 (ZEISS) at high vacuum condition. A layer of carbon powder is sprayed on the region of interest and then milled by a focused ion beam (ZEISS NVISION 40). The thickness of the membrane in the dry state is corrected from the measurement of the mass loss during the drying process to calculate the thickness in the wet state. The interfacial membrane was formed via complexation between 3.3\% of PFacid and 0.3\% chitosan. The error bars are standard deviation. }
	\end{figure}

\clearpage